\pdfoutput=1
\newcommand*{\ATLASLATEXPATH}{}
\documentclass[cernpreprint, texlive=2016, USenglish]{\ATLASLATEXPATH atlasdoc}
\usepackage[subfigure]{\ATLASLATEXPATH atlaspackage}
\usepackage{\ATLASLATEXPATH atlasbiblatex}
 
\usepackage{\ATLASLATEXPATH atlasphysics}
 
\graphicspath{{logos/}}
 
\usepackage{ANA-HDBS-2018-46-PAPER-defs}
 
\PreprintIdNumber{CERN-EP-2020-014}
\AtlasTitle{Search for
heavy Higgs bosons
decaying into two tau leptons with the ATLAS detector using $pp$ collisions
at $\sqrt{s}\,=\,\SI{13}{\text{TeV}}$
}

\AtlasAbstract{
A search for heavy neutral Higgs bosons is performed using
the LHC Run 2 data,
corresponding to an integrated luminosity of \SI{139}{fb$^{-1}$} of
proton--proton collisions at $\sqrt{s}\,=~\SI{13}{\text{TeV}}$ recorded with the ATLAS
detector. The search for heavy resonances is
performed over the mass range 0.2--\SI{2.5}{\text{TeV}} for the $\tau^+\tau^-$ decay
with at least one $\tau$-lepton decaying into final states
with hadrons.
The data are in good agreement with the background prediction of the Standard Model.
In the $M_{h}^{125}$ scenario of the Minimal Supersymmetric Standard Model,
values of
$\tan\beta\,>\,8$ and $\tan\beta\,>\,21$ are excluded at the 95\% confidence level for neutral
Higgs boson masses of
$\SI{1.0}{\text{TeV}}$ and $\SI{1.5}{\text{TeV}}$, respectively,
where $\tan\beta$ is the ratio of the vacuum expectation values of the two Higgs doublets.
}
 
\author{The ATLAS Collaboration}
 
\AtlasJournal{Phys.Rev.Lett.}
 
\AtlasJournalRef{Phys. Rev. Lett. 125 (2020) 051801}
\AtlasDOI{10.1103/PhysRevLett.125.051801}
\hypersetup{pdftitle={ATLAS document},pdfauthor={The ATLAS Collaboration}}
 
\begin{document}
 
\maketitle
 
The ATLAS and CMS collaborations discovered in 2012 a new boson with a mass of 125\,\GeV{}~\cite{ATLASHiggsJuly2012, CMSHiggsJuly2012}. Current measurements~\cite{Aad:2019mbh,Sirunyan:2018koj} indicate that the new particle is compatible with the Higgs boson predicted by the Standard Model (SM)~\cite{Englert:1964et,Higgs:1964pj,Guralnik:1964eu}. This discovery opens the way for studies of the structure of the Higgs sector.
Many theoretical models beyond the SM, such as two-Higgs-doublet models (2HDMs)~\cite{Branco:2011iw}, extend the Higgs sector to include a second Higgs doublet which implies the existence of new heavy pseudoscalar ($A$) and scalar ($H$) states, while the observed scalar particle would correspond to the lightest Higgs boson ($h$).
The decay probability of these scalar states into $\tau^{+}\tau^{-}$ pairs can be enhanced relative to other decay modes in 2HDMs of type II, such as the Minimal Supersymmetric SM
(MSSM)~\cite{Fayet:1976etFIX,Fayet:1977ycFIX}, the minimal extension of
the SM that realizes supersymmetry~\cite{Golfand:1971iwFIX,Volkov:1973ixFIX,Wess:1974twFIX,Salam:1974igFIX,Wess:1974jbFIX,Ferrara:1974puFIX}.
 
At tree level, the properties of the
MSSM Higgs sector depend only on two non-SM parameters, which can be
chosen to be the mass of the pseudoscalar Higgs boson, $m_A$, and the ratio of the
vacuum expectation values of the two Higgs doublets, $\tan\beta$. Beyond tree
level, the Higgs sector is affected by additional parameters, the choice of
which defines various MSSM benchmark scenarios. In the
$M^{125}_h$ scenario~\cite{MSSMBenchmarks2}, the parameters are such
that the mass of the lightest ${CP}$-even Higgs boson, $m_h$, is close to the measured
mass of the Higgs boson discovered at the LHC~\cite{LHC} and the masses of
all superparticles are heavy enough to only mildly affect the
production and decays of the MSSM Higgs bosons.
The couplings of the MSSM heavy Higgs bosons to down-type fermions are enhanced with
respect to the SM for large $\tanb$ values, resulting in increased branching
fractions to $\tau$-leptons and $b$-quarks, as well as a higher cross section
for Higgs boson production in association with $b$-quarks ($bbH$).
For the mass range considered in this Letter, the mass difference between the $A$ and $H$
bosons is much smaller than the experimental resolution and they are treated as degenerate in mass.
 
This Letter describes a search for massive scalar and pseudoscalar resonances decaying into a $\tau$-lepton pair.\footnote{Throughout this Letter the inclusion of charge-conjugate decay modes is implied.} The search is conducted on a sample of proton--proton collision data with an integrated luminosity of 139 fb$^{-1}$ at a center-of-mass energy of $\sqrt{s}\,=\,13\,\TeV$, collected with the ATLAS detector~\cite{PERF-2007-01,ATLAS-TDR-2010-19,PIX-2018-001} during the Run 2 of the LHC~(2015--2018)~\cite{Aad:2019hmz}. The $\tlep\thad$ and $\thad\thad$ decay channels are considered, where $\tlep$ denotes the decay of the $\tau$-lepton into neutrinos and an electron ($\taue$) or into neutrinos and a muon ($\taumu$) and  $\thad$ denotes the decay into a neutrino and hadrons.
This search improves on the results obtained by previous searches performed by the ATLAS and CMS collaborations at a center-of-mass energy of $\sqrt{s}\,=\,13\,\TeV$~\cite{HIGG-2015-08FIX,HIGG-2016-12,CMS-HIG-17-020} by about a factor of 4--5 for a scalar boson in the mass range 700--2500\,\GeV, thanks to improvements of the modeling of the top-quark background and of the backgrounds estimated from data, of the reconstruction of high-\pt{} $\tau$-leptons and the increase of integrated lumnosity.

The ATLAS detector at the LHC covers nearly the entire solid angle around the collision point.\footnote{ATLAS uses a right-handed coordinate system with its origin at the nominal interaction point (IP) in the center of the detector and the $z$-axis along the beam pipe. The $x$-axis points from the IP to the center of the LHC ring, and the $y$-axis points upward. Cylindrical coordinates $(r,\phi)$ are used in the transverse plane, $\phi$ being the azimuthal angle around the $z$-axis. The pseudorapidity is defined in terms of the polar angle $\theta$ as $\eta=-\ln\tan(\theta/2)$.}
It consists of an inner tracking detector surrounded by a thin superconducting solenoid, electromagnetic
and hadronic calorimeters, and a muon spectrometer incorporating three large superconducting
toroidal magnets.

Samples of Monte Carlo (MC) simulated events are used to optimize the event
selection,  estimate the signal efficiencies, and model some of the background processes.
The generators and parton showers used to simulate the different MC processes are summarized in Table~\ref{tab:generators}.
The production cross sections and branching fractions for the various MSSM
scenarios are calculated using procedures described in Refs.~\cite{Dittmaier:2011ti,LHCHXSWG2016}.  The cross sections for
gluon--gluon fusion (ggF) production  calculated with \SUSHI~\cite{Harlander:2012pb,Harlander:2016hcx} include next-to-leading-order (NLO)
supersymmetric-QCD corrections~\cite{Spira:1995rr,Harlander:2004tp,Harlander:2005rq,Degrassi:2010eu,Degrassi:2011vq,Degrassi:2012vt},
next-to-next-to-leading-order (NNLO) QCD corrections for the
top quark~\cite{Harlander:2002wh,Anastasiou:2002yz,Ravindran:2003um,Harlander:2002vv,Anastasiou:2002wq},
and light-quark electroweak effects~\cite{Aglietti:2004nj,
Bonciani:2010ms}.
The $bbH$ cross sections are calculated in the
five-flavor~\cite{Harlander:2003ai} and four-flavor schemes~\cite{Dittmaier:2003ej,Dawson:2003kb},
and the predictions are combined as described in
Refs.~\cite{Bonvini:2015pxa,Bonvini:2016fgf,Forte:2015hba,Forte:2016sja}.
The other production modes contribute negligibly in the $M^{125}_h$ scenario and are not considered.
The masses and mixing (and effective Yukawa couplings) of the
Higgs bosons are computed with \FEYNHIGGS~\cite{Heinemeyer:1998yj, Heinemeyer:1998np, Degrassi:2002fi, Frank:2006yh, Hahn:2013ria,
Williams:2011bu, Bahl:2016brp, Bahl:2017aev}.
Branching fractions of Higgs bosons are computed using a combination of results calculated by \FEYNHIGGS, \HDECAY~\cite{Djouadi:1997yw,Djouadi:2018xqq},
and PROPHECY4f~\cite{Bredenstein:2006rh, Bredenstein:2006ha}, following the procedure discussed in Ref.~\cite{Dittmaier:2011ti}.
The samples were
produced with the ATLAS simulation infrastructure~\cite{ATLASSIM} using
the full detector simulation performed by the {\GEANT}4~\cite{Agostinelli:2002hhFIX} toolkit,
with the exception of $bbH$ production of the MSSM
Higgs boson signal, for which the \AFII~\cite{ATLASSIM} fast
simulation framework was used.
 
\begin{table}[htb]
\caption{Generators used to describe the signal and background processes, parton distribution function (PDF) sets for the hard process, and models used for parton showering, hadronization and the underlying event (UEPS). The orders of the total cross sections used to normalize the events are also given. $V$ represents either $W$ or $Z$ gauge bosons.}
\centering
\setlength{\extrarowheight}{0pt}
\scriptsize
\begin{tabular}{lllll}
\toprule
Process & Generator & PDF & UEPS & Cross-section order \\ \midrule
ggF & \POWHEGBOXV{v2} \cite{Nason:2004rx,Frixione:2007vw,Alioli:2010xd,POWHEG,Powheg_4} & CT10~\cite{Lai:2010vv} & \PYTHIAV{8.1}~\cite{Sjostrand:2007gs} & See text \\
$bbH$ & MG5\_aMC@NLO 2.1.2~\cite{Alwall:2014hca,Wiesemann:2014ioa} &  CT10 & \PYTHIAV{8.2}~\cite{Sjostrand:2014zea} & See text \\
$W$+jets & \SHERPAV{2.2.1}~\cite{Bothmann:2019yzt} & NNPDF 3.0 NNLO~\cite{Ball:2014uwaFIX} & \SHERPAV{2.2.1}~\cite{Schumann:2007mgFIX} & NNLO~\cite{Anastasiou:2003ds} \\
$Z$+jets & \POWHEGBOXV{v1}~\cite{Nason:2004rx,Frixione:2007vw,Alioli:2010xd,Alioli:2008gx} & CT10 & \PYTHIAV{8.1} & NNLO~\cite{Anastasiou:2003ds} \\
$VV$/$V\gamma^*$ & \SHERPAV{2.2} & NNPDF 3.0 NNLO & \SHERPAV{2.2} & NLO \\
$t\bar{t}$ & \POWHEGBOXV{v2}~\cite{Frixione:2007nw,Nason:2004rx,Frixione:2007vw,Alioli:2010xd} & NNPDF 3.0 NLO &  \PYTHIAV{8.2} & NNLO+NNLL~\cite{Beneke:2011mq,Cacciari:2011hy,Baernreuther:2012ws,Czakon:2012zr,Czakon:2012pz,Czakon:2013goa,Czakon:2011xxFIX}\\
Single $t$ & \POWHEGBOXV{v2} \cite{Re:2010bp,Frederix:2012dh,Alioli:2009je,Nason:2004rx,Frixione:2007vw,Alioli:2010xd} & NNPDF 3.0 NLO & \PYTHIAV{8.2} & NNLO+NNLL \cite{Aliev:2010zk,Kant:2014oha} \\ \bottomrule
\end{tabular}
\label{tab:generators}
\end{table}
\par
In this search, the leptonic $\tau$ decays are identified by their charged decay product, either an electron or a muon.
Electron candidates are reconstructed from energy deposits
in the electromagnetic calorimeter
associated with a charged-particle track measured in the inner detector~\cite{Aad:2019tso}.
They are required  to have $|\eta| < 2.47$.
The transition region between the barrel and endcap calorimeters ($1.37 < |\eta| < 1.52$) is excluded.
 
Muon candidates are reconstructed in the range $|\eta| < 2.5$ by matching tracks
found in the muon spectrometer to tracks found in the inner detector~\cite{PERF-2015-10}.
The selected leptons in the $\lephad$ channel are required to have a transverse momentum $\pt > 30\,\GeV$,
pass the ``medium'' quality requirement for both the electrons~\cite{Aad:2019tso} and muons~\cite{PERF-2015-10} and satisfy a $\pt$- and $\eta$-dependent isolation criterion called ``Gradient'', which uses calorimetric and tracking information. The  efficiencies for the identification and isolation criteria are given in Refs.~\cite{Aad:2019tso,PERF-2015-10}.
\par
Jets are reconstructed from topological clusters~\cite{PERF-2014-07} of energy depositions
in the calorimeter using the \antikt algorithm~\cite{Cacciari:2008gp}, with a radius parameter value $R = 0.4$~\cite{Fastjet}.
The average energy contribution from pileup is subtracted according to the jet area and
the jets are calibrated as described in Ref.~\cite{PERF-2016-04}.
Jets are required to have $\pt > \SI{20}{\GeV}$ and $|\eta| < 2.5$.
The effect of pileup is reduced by using tracking information associated with the calorimeter-based jets to reject those not originating from the primary vertex~\cite{PERF-2014-03}. The primary vertex is chosen as the proton--proton vertex candidate with the highest sum of the squared transverse momenta of the associated tracks.
 
In order to identify jets containing $b$-hadrons~($b$-jets), a multivariate algorithm (MV2) is used~\cite{Aad:2019aic}.
The algorithm has an average efficiency of 70\% for $b$-jets and rejections of approximately 9, 36, and 300 for $c$-jets, $\tau$ decays with hadrons, and jets initiated by light quarks or gluons, respectively,
as determined in simulated $t\bar{t}$ events. Correction factors are applied to the simulated event samples to compensate for differences between data and simulation in the $b$-tagging efficiencies for $b$-jets, $c$-jets and light-flavor jets.
\par
Hadronic $\tau$ decays are composed of a neutrino and a set of visible decay products~(\tauhadvis),
typically one or three charged pions and up to two neutral pions.
The \tauhadvis candidates reconstructed from seeding jets~\cite{ATLAS-CONF-2017-029} must have $\pt > 25$~($65$)\,$\GeV$ in the \lephad~(\hadhad) channel,
$|\eta| < 2.5$ excluding $1.37 < |\eta| < 1.52$, one or three associated tracks and an electric charge of $\pm 1$.
A boosted-decision-tree identification procedure, based on calorimetric shower shapes and
tracking information, is used to reject jets.
 
The \tauhadvis candidates must satisfy ``loose'' or ``medium'' $\tau$ identification criteria~\cite{ATLAS-CONF-2017-029} with efficiencies of about 85\%~(75\%) and 75\%~(60\%) for one-track~(three-track) \tauhadvis candidates, respectively.
The rejections factors of  ``loose'' and ``medium'' $\tau$ identification in multijet events are about 20~(200) and 30~(500) for one-track~(three-track) \tauhadvis{} candidates, respectively.
\par
The missing transverse momentum, \metvec, is calculated as the negative vectorial sum of the \ptvec
of all fully reconstructed and calibrated physics objects~\cite{ATLAS-CONF-2018-023}.
In addition, this procedure includes a soft term, which is calculated using the inner-detector tracks that
originate from the hard-scattering vertex but are not associated with reconstructed objects.
\par
Events in the \lephad channel are selected using single-electron and single-muon  triggers
with \pt thresholds ranging from 20 to 26\,\GeV{} and various isolation criteria~\cite{TRIG-2018-05FIX,TRIG-2016-01}.
The events must contain at least one \tauhadvis candidate passing the medium identification and
exactly one isolated lepton~($\ell$).
The \tauhadvis candidate must have $|\eta| < 2.3$ to reduce misidentified-electron background~\cite{Aad:2014rga}.
The isolated lepton and the \tauhadvis candidate must have opposite electric charge
and be back to back in the transverse plane: $|\Delta\phi (\ptvec^\ell,\ptvec^{\tauhadvis})|>2.4$ rad.
To reduce background from $W$+jets production the transverse mass
$m_\text{T}(\ptvec^\ell,\metvec)=\sqrt{2\pt^\ell\met [1-\cos\Delta\phi(\ptvec^\ell,\metvec)]}$, calculated with the lepton $\pt$ and the event $\metvec$, must be less than 40\,\GeV.
To reduce background from $Z \to ee$ production in the $\taue\tauhad$ channel, events
in which
the isolated lepton
and the \tauhadvis candidate have an invariant mass between 80\,\GeV{} and 110\,\GeV{} are rejected. The background contribution from  $Z \to \mu\mu$ in the  $\taumu\tauhad$ channel is found to be negligible.
The signal acceptance times efficiency for each of the $\taue\tauhad$ and $\taumu\tauhad$ channels varies between 2\% and 7\% for signals with masses of 0.2--2.5\,\TeV.\footnote{The acceptance is calculated with respect to the sum of all $\tau$ decay modes. The efficiency is calculated taking into account detector acceptance, reconstruction and selection efficiencies.}
\par
Events in the \hadhad channel are selected by single-$\tau$ triggers with \pt thresholds of 80\,\GeV{} (5.4\,fb$^{-1}$ from June 2015 to May 2016), 125\,\GeV{} (9.3\,fb$^{-1}$ in May--June 2016) and 160\,\GeV{} (124\,fb$^{-1}$ from June 2016 to October 2018).
Events must contain at least two \tauhadvis candidates
and no electrons or muons.
The \pt{} of the leading \tauhadvis candidate
must exceed the trigger \pt threshold by 5\,\GeV.
The leading (subleading) \tauhadvis candidate must satisfy the medium~(loose) identification criteria.
The two  \tauhadvis must have opposite electric charge and be back to back in the transverse plane: $|\Delta\phi (\ptvec^{\tauhadvis^{1}},\ptvec^{\tauhadvis^{2}})|>2.7$~rad.
The signal acceptance times efficiency varies between 2\% and 20\% for signals with masses of 0.35--2.5\,\TeV, and it decreases rapidly for lower mass values due to the selection criteria imposed on the $\pt$ of the decay products of the $\tau$-leptons.
\par
Events satisfying the selection criteria of either channel are divided into categories to exploit the different production modes in the MSSM:
the $b$-tag category for events containing at least one $b$-jet and the $b$-veto category for events containing no $b$-jets. These categories are the signal regions used by the analysis.
\par
The $\tau\tau$ mass reconstruction is crucial for good separation between signal and background events.
However, its reconstruction is challenging due to the presence of neutrinos from the $\tau$-lepton decays.
The mass reconstruction used for both channels is the total transverse mass, defined as
$m^\text{tot}_\text{T}\,=\,\sqrt{(\pt^{\tau_{1}}+\pt^{\tau_{2}}+\met)^2-(\ptvec^{\tau_{1}}+\ptvec^{\tau_{2}}+\metvec)^2}$
for either ($\ell$,$\tauhadvis$) or ($\tauhadvis^{1}$,$\tauhadvis^{2}$) as ($\tau_{1}, \tau_{2}$).
\par
The dominant background contribution in the \lephad channel arises from processes
where the \tauhadvis candidate originates from a jet.
Such background events are divided into those where the selected lepton is correctly identified,
mainly from $W$+jets~($t\bar{t}$) production in the $b$-veto~($b$-tag) category, and
those where the selected lepton arises from a jet, mainly from multijet production.
These contributions are estimated using a data-driven technique, which is similar to that described in Ref.~\cite{HIGG-2016-12}.
Three orthogonal control regions are defined using the same selection as for the signal region,
except that the lepton candidate fails isolation requirements in CR-0,
the \tauhadvis candidate fails $\tau$ identification in CR-1, and both fail these conditions in CR-2.
The multijet background events are estimated from CR-0 weighted with lepton correction factors, called fake-factors, which are ratios of the numbers of lepton candidates passing and failing the isolation requirements~\cite{HIGG-2016-12}.\footnote{Hereafter, fake-factors refer to ratios of the number of candidates passing a certain identification requirement to the number of candidates failing the requirement.}
The $W$+jets~($t\bar{t}$) background events are estimated from CR-1  after subtracting the multijet background contributions estimated from CR-2 corrected with lepton fake-factors.
Real $\tau$-lepton contributions in CR-1 are subtracted using MC simulation.
The $\tau$-lepton fake-factor weights measured in data are then applied to the events in CR-1 to estimate the $W$+jets~($t\bar{t}$) background in the signal region.
Backgrounds where both the lepton and \tauhadvis candidates originate from electrons, muons or $\tau$-leptons
arise from $Z/\gamma^* \to \tau\tau$ production in the $b$-veto category and $t\bar{t}$ production in the
$b$-tag category, with minor contributions from $Z/\gamma^* \to \ell\ell$, diboson and single top-quark production.
These contributions are estimated using MC simulation.
To constrain the normalization of the $t\bar{t}$ contribution, a top-quark control region enhanced in $t\bar{t}$ events
is defined
by substituting the transverse mass requirement with $m_\text{T}(\ptvec^\ell,\metvec)>110$~($100$)~\GeV{} in the $b$-tag category of the  $\taue\tauhad$~($\taumu\tauhad$) channel.
This region is included in the fitting procedure. Other major background contributions can be adequately constrained in the signal regions.
\par
The dominant background contribution in the \hadhad channel is from multijet production,
which is estimated using a data-driven technique described in Ref.~\cite{HIGG-2016-12}:
the background is estimated from a control region whose events pass the same selection as for the signal region, except the subleading \tauhadvis candidates fail $\tau$ identification.
Then the events are weighted with fake-factors measured in a region enriched with multijet events to obtain the multijet background estimation in the signal region.
The other non-negligible
backgrounds contributions are $Z/\gamma^* \to \tau\tau$ production
in the $b$-veto category, $t\bar{t}$ production in the $b$-tag category, and to a lesser extent $W$~($\to \tau\nu, \ell\nu$)+jets,
single top-quark, diboson and $Z/\gamma^*$~($\to \ell\ell$)+jets production.
These contributions are estimated using MC simulation.
To improve the modeling of jets faking hadronic $\tau$ decays~(fake $\tau$-leptons), events in the simulation that contain
quark- or gluon-initiated jets that are misidentified as \tauhadvis candidates
are corrected to follow rates of fake $\tau$-leptons measured in $W$+jets and $t\bar{t}$ enhanced regions in data.
\par
Uncertainties affecting the simulated signal and background contributions are considered in the statistical analysis.
These include uncertainties associated with the determination of the integrated luminosity~\cite{LUCID2,ATLAS-CONF-2019-021}, the detector
simulation, the theoretical cross sections, and the background modeling.
For MSSM Higgs boson samples,
various sources of uncertainty which affect the signal acceptance are considered, such as the impact of varying the
factorization and renormalization scales and uncertainties in the modeling of initial- and final-state radiation,
as well as multiple parton interactions.
The sensitivity of the search is limited by statistical uncertainties, especially for scalars with mass values above \SI{600}{\GeV}.
The main systematic uncertainties are shown in Table~\ref{tab:systematics}. They are related to the determination of the \tauhadvis{} identification efficiency and energy scale, estimation of the backgrounds with misidentified \tauhadvis{} and modeling of $Z$+jets background.
The uncertainty in the \tauhadvis{} identification
efficiency is determined from measurements of $Z\to\tau\tau$ events and, for the high \pt{} regime, an additional uncertainty is assigned from the validation of the  \tauhadvis{} properties in high-\pt{} dijet events.
The uncertainty in the \tauhadvis{} energy scale is derived from $Z\to\tau\tau$ events as well, and from single hadron test-beam data, and it is validated for high-\pt{} $\tauhadvis$ with top-quark events and $Z$($\to\tau\tau$) events with large transverse momentum.
Uncertainties in the determination of backgrounds with  misidentified \tauhadvis{} include the uncertainty from the subtraction of other backgrounds in the control regions, the uncertainty from the limited number of events in the control regions and the uncertainty from differences in the jet composition between control regions and signal regions. For $Z$+jets production,  cross-section and modeling uncertainties
are taken from Refs.~\cite{EXOT-2015-07,Andersen:2014efa}.

\begin{table}[htb]
 
\caption{Relative increase in the expected 95\% CL upper limits for the production cross section times branching fraction relative to the statistical only expected limit for each systematic uncertainty under consideration, shown for scalar bosons with mass of 400\,\GeV~and 1\,\TeV~produced via ggF and $bbH$ production.
}
\centering
\setlength{\extrarowheight}{0pt}
\scriptsize
 
\begin{tabular}{lllll}
\toprule
Source  & ggF~(400\,\GeV) & ggF~(1\,\TeV) & $bbH$~(400\,\GeV) & $bbH$~(1\,\TeV) \\ \midrule
Tau id. efficiency        & 0.14 & 0.16    & 0.12 & 0.08 \\
Tau energy scale          & 0.33 & 0.09    & 0.22 & 0.03 \\
$Z$+jets bkg. modeling    & 0.27 & 0.19    & 0.08 & 0.04 \\
Mis-id. \tauhadvis{} bkg. & 0.22 & 0.01    & 0.14 & 0.03 \\
Others                    & 0.09 & 0.04    & 0.11 & 0.02 \\ \midrule
Total                     & 0.54 & 0.28    & 0.45 & 0.13 \\ \bottomrule
\end{tabular}
\label{tab:systematics}
\end{table}
\FloatBarrier
\par
A simultaneous fit of the \mttot distributions of the top-quark control region and of the $b$-veto and $b$-tag categories of the \lephad and \hadhad channels is performed in the statistical analysis.
The numbers of observed events in the $b$-veto and $b$-tag categories of the \lephad channel are 728,174 and 19,542,
while event yields of 728,200$\pm$2900 and 19,600$\pm$400 for the background-only hypothesis are obtained from the statistical analysis, which includes the fit of the nuisance parameters associated with the systematic uncertainties.
 
\begin{figure}[htb]
\begin{center}
\subfigure[]{\includegraphics[width=0.47\linewidth]{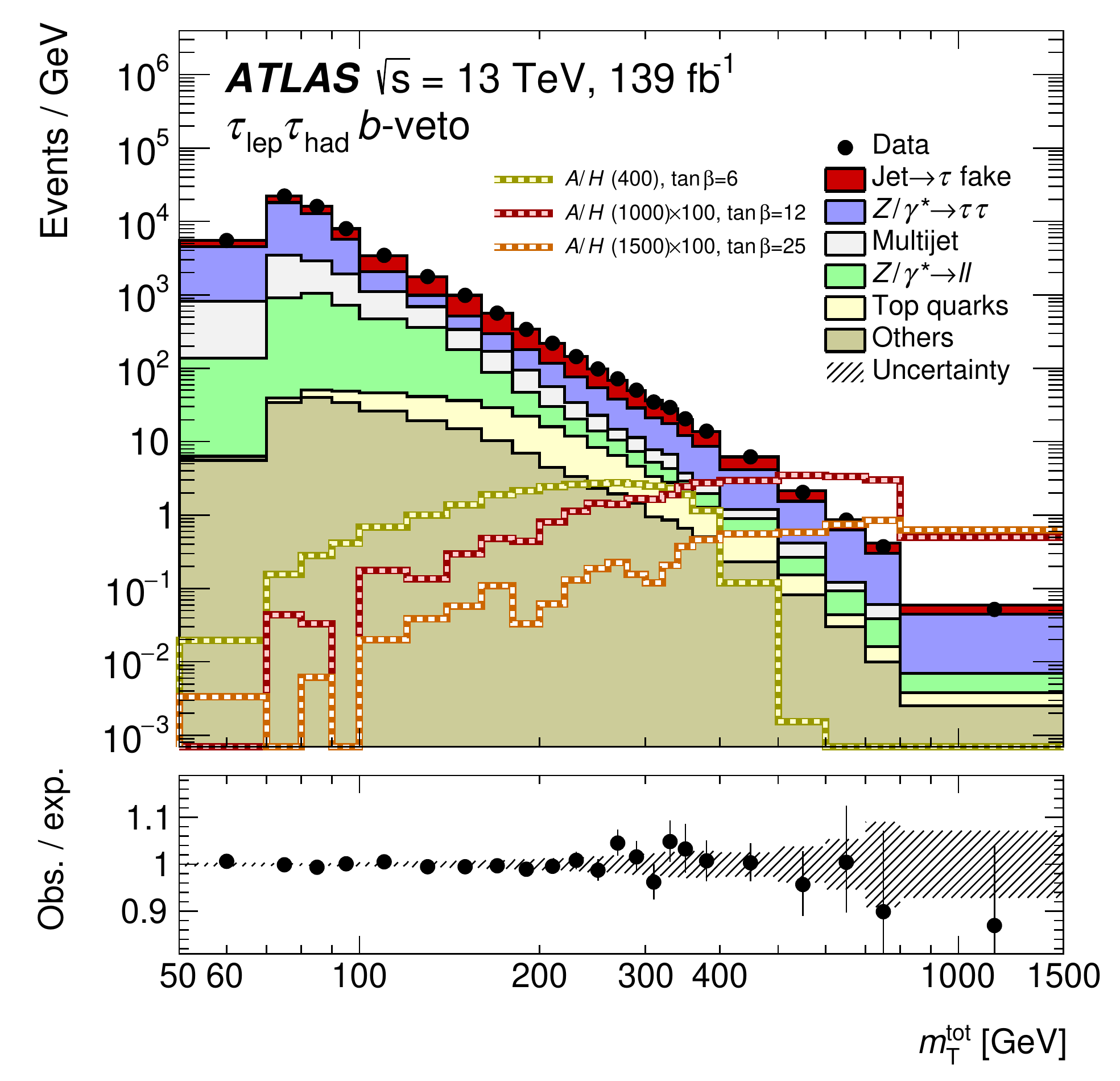}}
\subfigure[]{\includegraphics[width=0.47\linewidth]{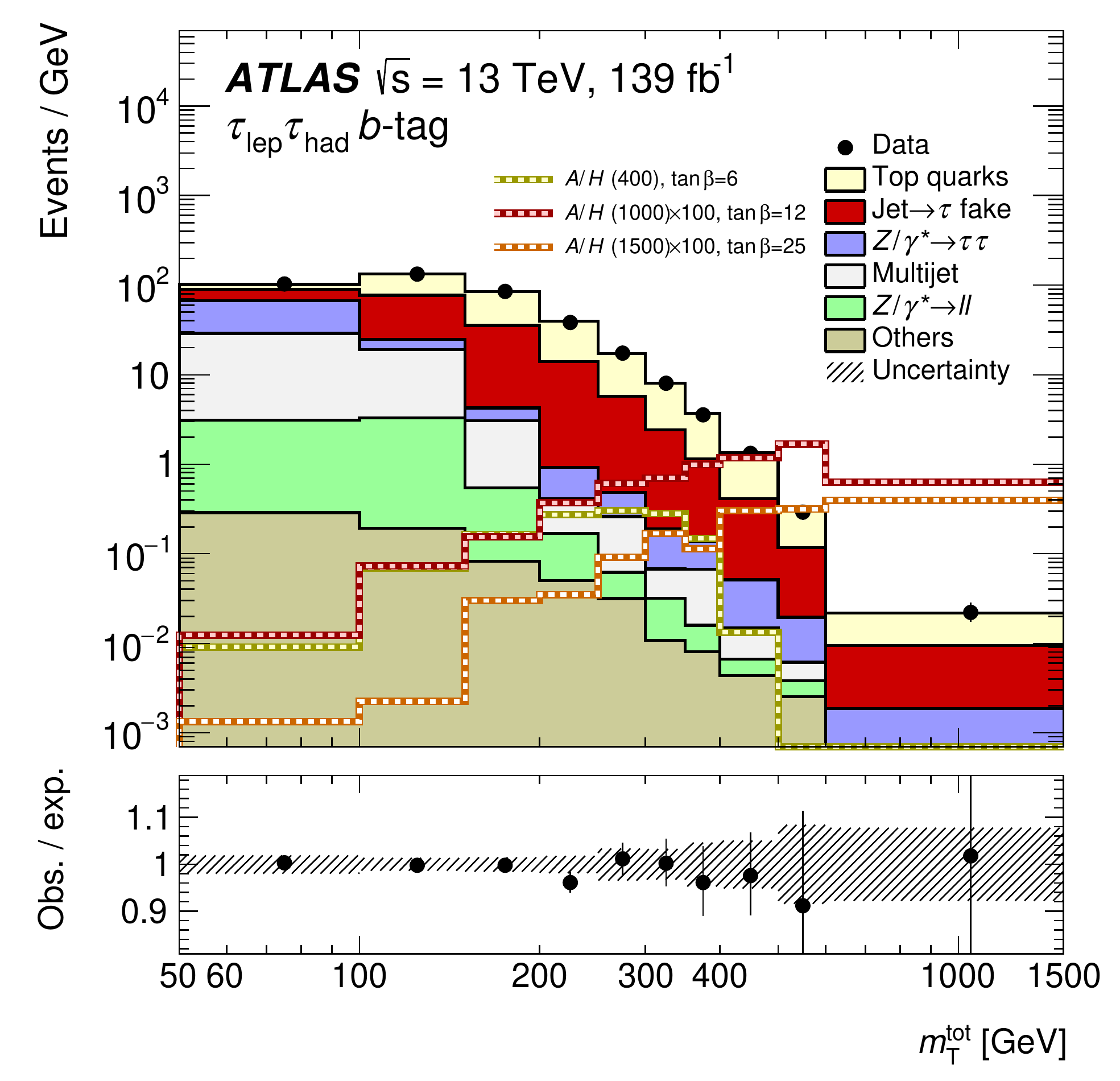}}\\
\subfigure[]{\includegraphics[width=0.47\linewidth]{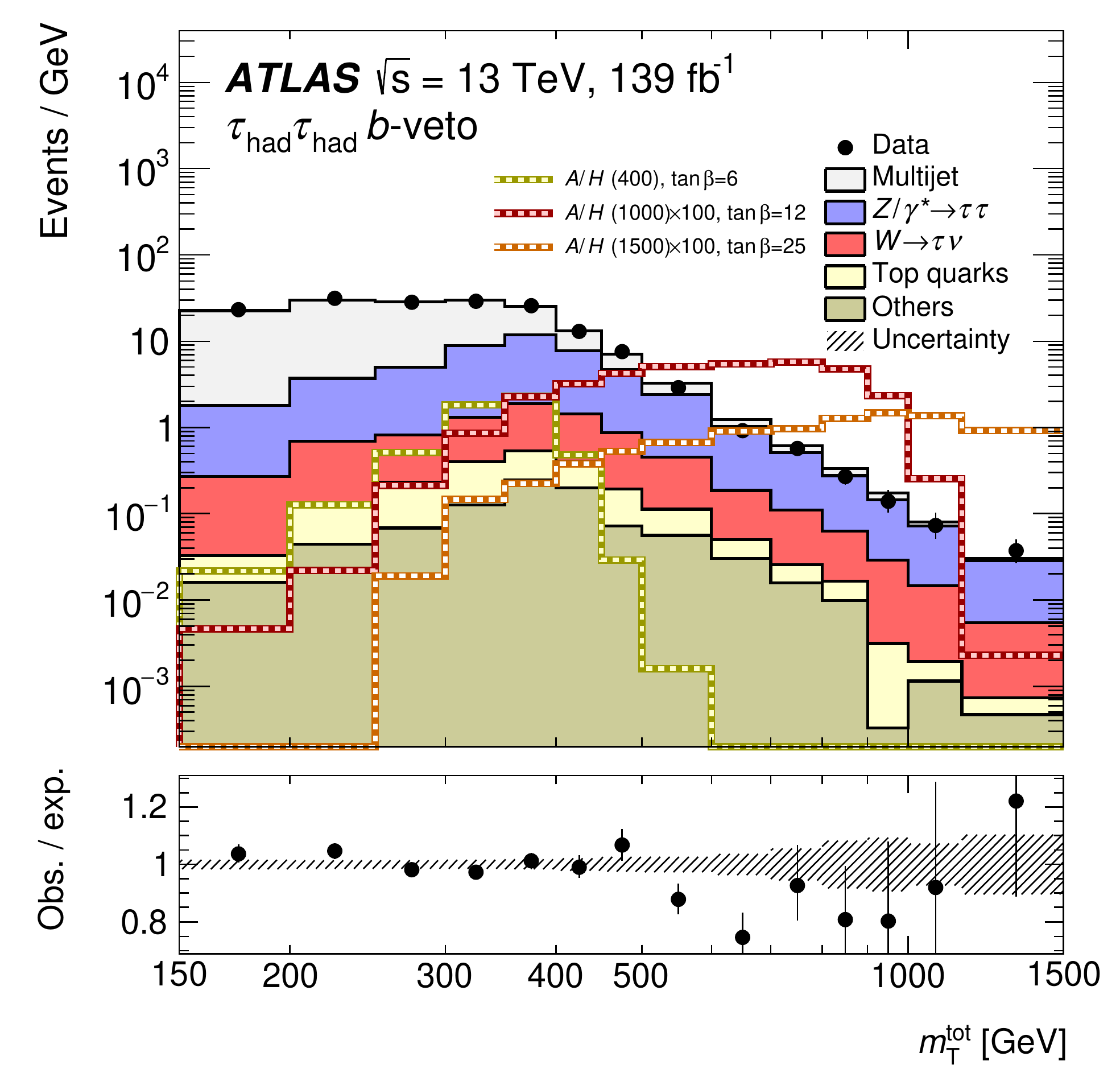}}
\subfigure[]{\includegraphics[width=0.47\linewidth]{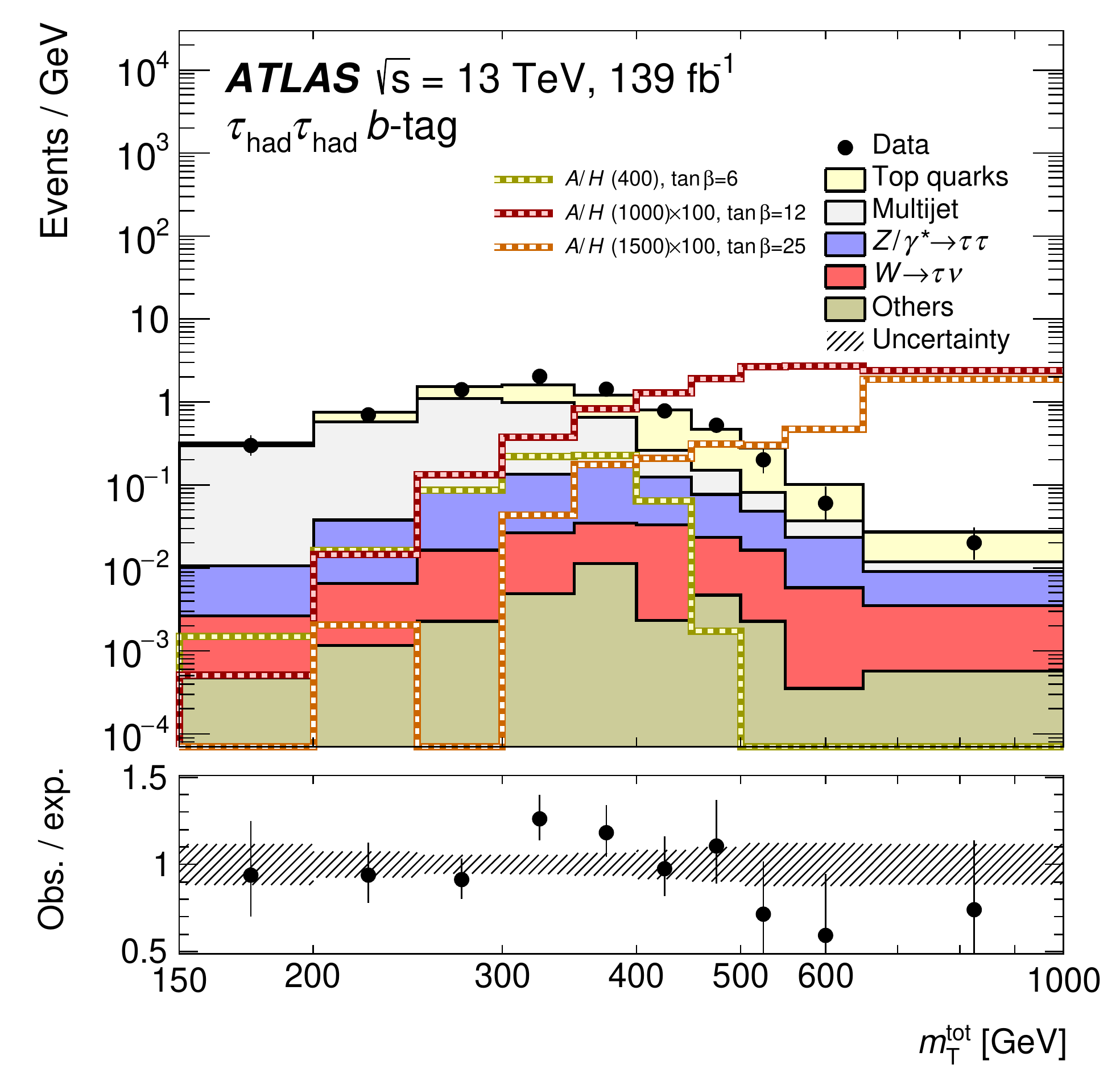}}
\end{center}
\caption{
The \mttot for the $b$-veto (left) and $b$-tag (right) categories of the \lephad channel (top) and \hadhad channel (bottom).
The binning displayed is that entering into the fit.
The predictions and uncertainties for the background
processes are obtained from the fit
assuming the background-only hypothesis.
Expectations from signal processes are superimposed.
Overflows are included in the last bin of the distributions.
}
\label{fig:mTtot}
\end{figure}

For the \hadhad channel, the numbers of observed events in the $b$-veto and $b$-tag categories
are 8420 and 381, and the fitted event yields
from background processes are 8430$\pm$150 and 368$\pm$27.
The \mttot distributions obtained from the fit performed simultaneously in the $b$-veto and $b$-tag categories of the two channels
are shown in Figure~\ref{fig:mTtot}.
\par
\begin{figure}[htb]
\begin{center}
\subfigure[]{\includegraphics[width=0.31\textwidth]{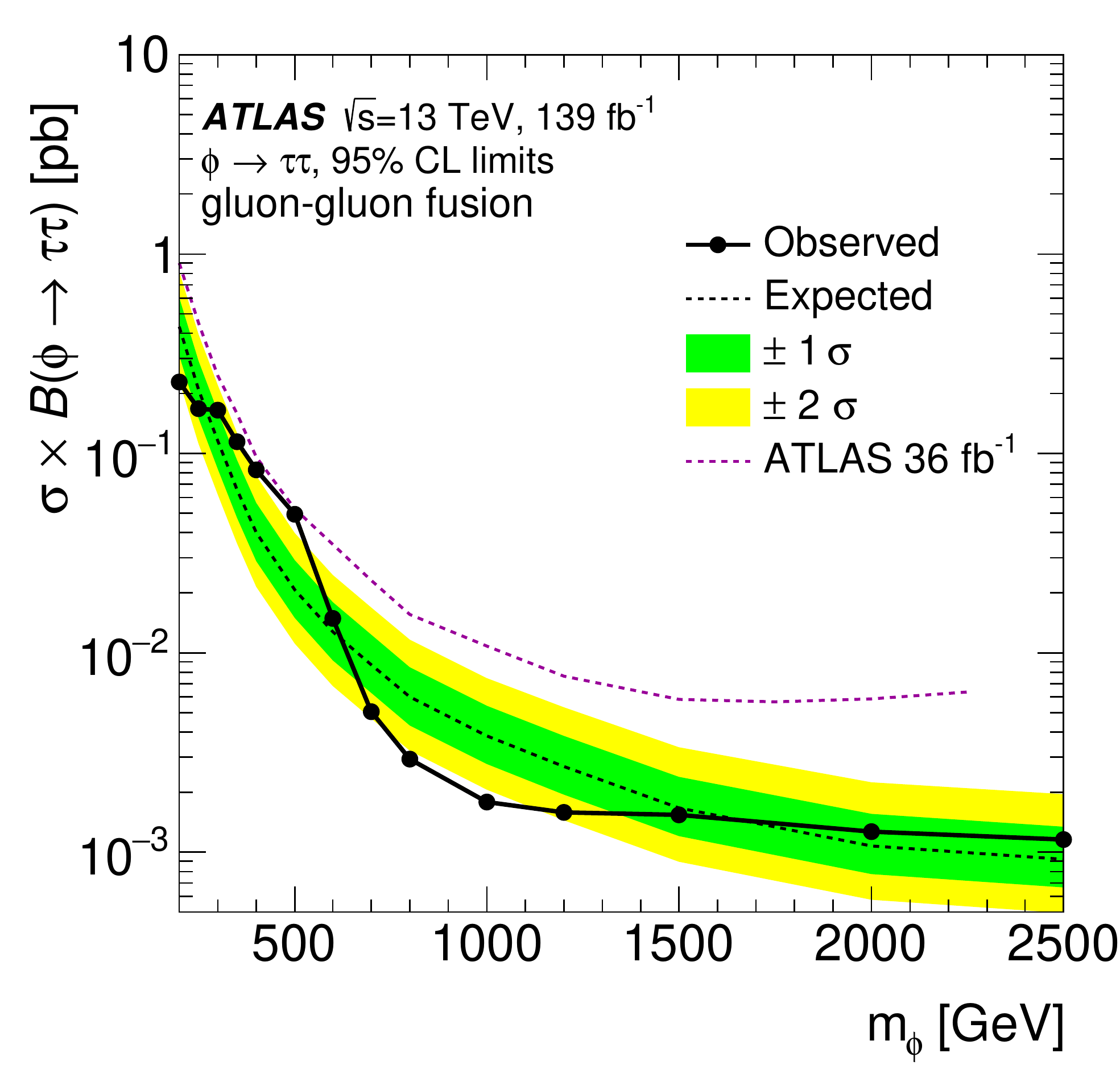}\label{fig:limits-xsec1}}
\subfigure[]{\includegraphics[width=0.31\textwidth]{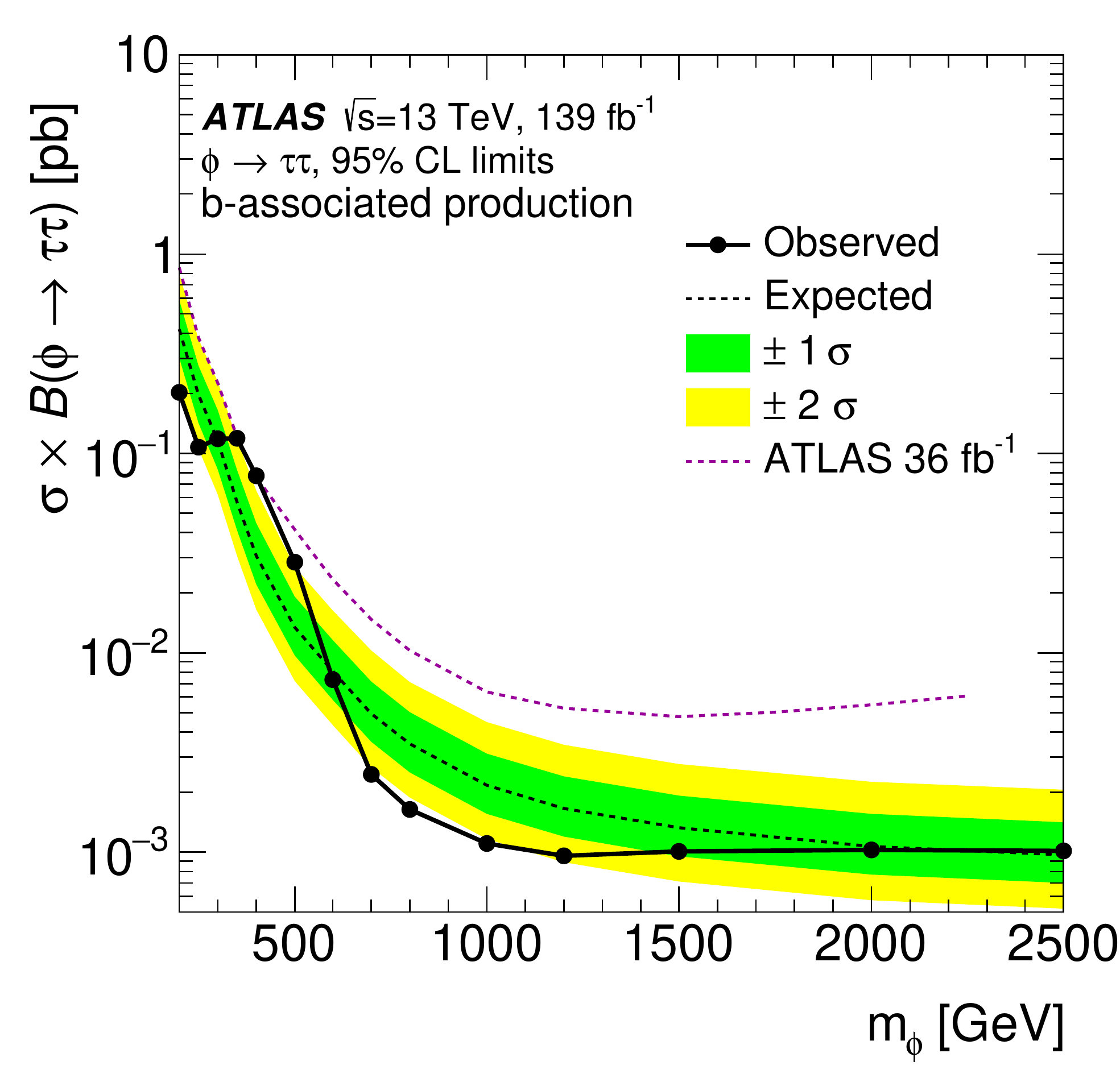}\label{fig:limits-xsec2}}
\subfigure[]{\includegraphics[width=0.31\textwidth]{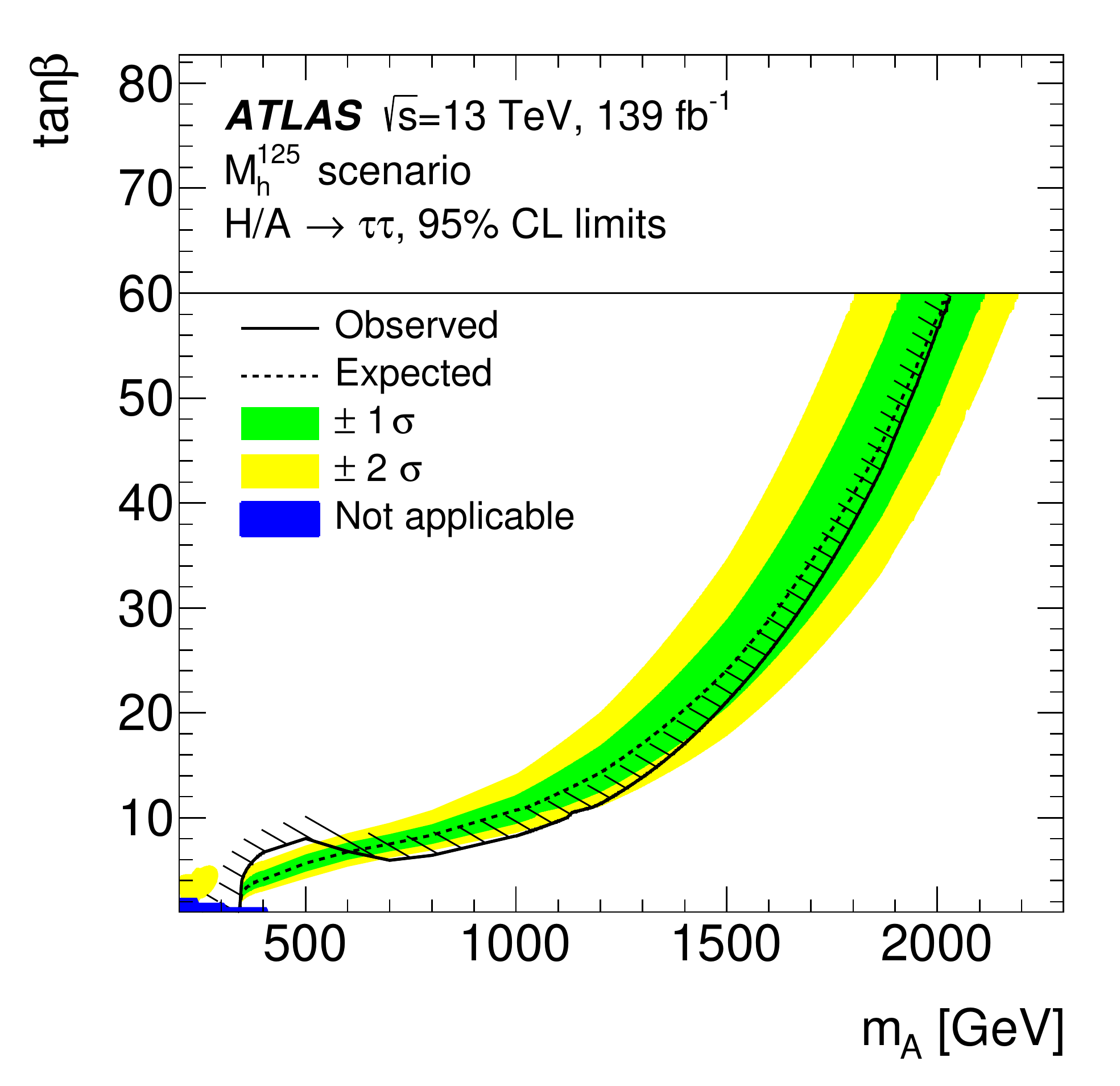}\label{fig:limits-xsec3}}
\end{center}
\caption{
The observed and expected 95\% CL upper limits on the production
cross section times branching fraction for a scalar
boson ($\phi$) produced via (a) ggF and (b) $b$-associated production.
The limits are calculated
from a statistical combination of the \lephad and \hadhad channels.
The excluded region from the 2015--2016 data ATLAS search~\cite{HIGG-2016-12}
is depicted by the dotted pink line.
The 95\% CL upper limits on $\tan\beta$ as a function of $m_A$ in the $M^{125}_h$ scenario is shown~(c).
The lowest value of $\tan\beta$ considered for the  $M^{125}_h$ scenario is 0.5.
In the small lower-left region shown in solid blue, the mass splitting between $A$ and $H$ bosons
is above 50\% of the mass resolution and therefore the simple addition of the cross sections is
not valid. However, this region of parameter space in
the $M^{125}_h$ scenario
provides predictions that are incompatible with the measured
mass value of the observed Higgs boson by more than $3\,\sigma$.
The exclusion limit around  $m_A=$~350\,\GeV{} reflects the behavior of the $A\to\tau\tau$
branching fraction close to the $A\to \ttbar$ kinematic threshold for low $\tanb$.
The hatched area defines which side of the curve is excluded by the search.
}
\end{figure}

The data are found to be in good agreement with the obtained background yields, and
the results are given in terms of exclusion limits.
Upper limits on the cross section times branching fraction for  a scalar
boson (generically called $\phi$) decaying into $\tau$-lepton pairs are set at the 95\% confidence level (CL) as a function of the boson mass. They are computed using a modified frequentist $\text{CL}_\text{S}$ method~\cite{Read:2002hqFIX} with the profile likelihood ratio as the test statistic. The asymptotic approximation is used~\cite{Cowan:2010jsFIX}.
The upper limits cover the mass range 0.2--\SI{2.5}{\TeV}
and are shown for a production entirely via ggF  in
Figure~\ref{fig:limits-xsec1} and  entirely via $b$-quark associated production in Figure~\ref{fig:limits-xsec2}.
The observed (expected) upper limits are \SI{1.8}{fb}~(\SI{3.8}{fb}) for ggF and \SI{1.1}{fb}~(\SI{2.2}{fb}) for $bbH$ production at $m_{\phi}=1$~\TeV. For  ggF, the lowest local $p_0$, the probability that the background can produce a fluctuation greater than the excess observed in data, is 0.014 (2.2\,$\sigma$) at $m_\phi=$~400\,\GeV, while for $bbH$ production it is 0.003 (2.7\,$\sigma$) at $m_\phi=$~400\,\GeV. The natural width of the scalar boson is assumed to be negligible compared to the experimental resolution.
Results are interpreted in terms of the MSSM in Figure~\ref{fig:limits-xsec3}, which shows the regions in the $m_A$--$\tan\beta$ plane excluded at the 95\% CL in the $M^{125}_h$ scenario.
The observed~(expected) upper limits exclude $\tan\beta >21$~(24) for $m_A=1.5$~\TeV.
 
\FloatBarrier
\par
In conclusion, a search for heavy neutral Higgs bosons decaying into a pair of $\tau$-leptons is performed in the mass range 0.2--\SI{2.5}{\TeV} using a data sample
corresponding to an integrated luminosity of \SI{139}{fb$^{-1}$} from
proton--proton collisions at $\sqrt{s}\,=\,\SI{13}{\TeV}$ recorded by the ATLAS detector at the LHC.
No significant
excess over the expected SM backgrounds is found.
Upper limits on the cross section for the production of a scalar boson times the branching fraction to $\tau\tau$ final states are set at the 95\% CL,
significantly increasing the sensitivity and explored mass range compared to previous searches.
They are in the range 240--\SI{1.2}{fb} (230--\SI{1.0}{fb}) for gluon--gluon fusion ($b$-associated) production of scalar bosons with masses
of 0.2--\SI{2.5}{\TeV}.
In the $M^{125}_h$ scenario, the data exclude
$\tanb > 8$ for $\MA =\SI{1.0}{\TeV}$ and
$\tanb> 21$ for $\MA = \SI{1.5}{\TeV}$ at the 95\% CL.
 
We thank CERN for the very successful operation of the LHC, as well as the
support staff from our institutions without whom ATLAS could not be
operated efficiently.
 
We acknowledge the support of ANPCyT, Argentina; YerPhI, Armenia; ARC, Australia; BMWFW and FWF, Austria; ANAS, Azerbaijan; SSTC, Belarus; CNPq and FAPESP, Brazil; NSERC, NRC and CFI, Canada; CERN; CONICYT, Chile; CAS, MOST and NSFC, China; COLCIENCIAS, Colombia; MSMT CR, MPO CR and VSC CR, Czech Republic; DNRF and DNSRC, Denmark; IN2P3-CNRS and CEA-DRF/IRFU, France; SRNSFG, Georgia; BMBF, HGF and MPG, Germany; GSRT, Greece; RGC and Hong Kong SAR, China; ISF and Benoziyo Center, Israel; INFN, Italy; MEXT and JSPS, Japan; CNRST, Morocco; NWO, Netherlands; RCN, Norway; MNiSW and NCN, Poland; FCT, Portugal; MNE/IFA, Romania; MES of Russia and NRC KI, Russia Federation; JINR; MESTD, Serbia; MSSR, Slovakia; ARRS and MIZ\v{S}, Slovenia; DST/NRF, South Africa; MINECO, Spain; SRC and Wallenberg Foundation, Sweden; SERI, SNSF and Cantons of Bern and Geneva, Switzerland; MOST, Taiwan; TAEK, Turkey; STFC, United Kingdom; DOE and NSF, United States of America. In addition, individual groups and members have received support from BCKDF, CANARIE, Compute Canada and CRC, Canada; ERC, ERDF, Horizon 2020, Marie Sk{\l}odowska-Curie Actions and COST, European Union; Investissements d'Avenir Labex, Investissements d'Avenir Idex and ANR, France; DFG and AvH Foundation, Germany; Herakleitos, Thales and Aristeia programmes co-financed by EU-ESF and the Greek NSRF, Greece; BSF-NSF and GIF, Israel; CERCA Programme Generalitat de Catalunya and PROMETEO Programme Generalitat Valenciana, Spain; G\"{o}ran Gustafssons Stiftelse, Sweden; The Royal Society and Leverhulme Trust, United Kingdom.
 
The crucial computing support from all WLCG partners is acknowledged gratefully, in particular from CERN, the ATLAS Tier-1 facilities at TRIUMF (Canada), NDGF (Denmark, Norway, Sweden), CC-IN2P3 (France), KIT/GridKA (Germany), INFN-CNAF (Italy), NL-T1 (Netherlands), PIC (Spain), ASGC (Taiwan), RAL (UK) and BNL (USA), the Tier-2 facilities worldwide and large non-WLCG resource providers. Major contributors of computing resources are listed in Ref.~\cite{ATL-GEN-PUB-2016-002}.

\clearpage
\appendix
 
\printbibliography
 
\clearpage 
 
\begin{flushleft}
\hypersetup{urlcolor=black}
{\Large The ATLAS Collaboration}

\bigskip

\AtlasOrcid[0000-0002-6665-4934]{G.~Aad}$^\textrm{\scriptsize 102}$,    
\AtlasOrcid[0000-0002-5888-2734]{B.~Abbott}$^\textrm{\scriptsize 128}$,    
\AtlasOrcid{D.C.~Abbott}$^\textrm{\scriptsize 103}$,    
\AtlasOrcid[0000-0002-2788-3822]{A.~Abed~Abud}$^\textrm{\scriptsize 36}$,    
\AtlasOrcid[0000-0002-1002-1652]{K.~Abeling}$^\textrm{\scriptsize 53}$,    
\AtlasOrcid[0000-0002-2987-4006]{D.K.~Abhayasinghe}$^\textrm{\scriptsize 94}$,    
\AtlasOrcid[0000-0002-8496-9294]{S.H.~Abidi}$^\textrm{\scriptsize 166}$,    
\AtlasOrcid[0000-0002-8279-9324]{O.S.~AbouZeid}$^\textrm{\scriptsize 40}$,    
\AtlasOrcid{N.L.~Abraham}$^\textrm{\scriptsize 155}$,    
\AtlasOrcid[0000-0001-5329-6640]{H.~Abramowicz}$^\textrm{\scriptsize 160}$
% ...
\bigskip
\\

% ...
$^{36}$CERN, Geneva; Switzerland.\\
% ...
$^{40}$Niels Bohr Institute, University of Copenhagen, Copenhagen; Denmark.\\
% ...
$^{53}$II. Physikalisches Institut, Georg-August-Universit\"{a}t G\"ottingen, G\"ottingen; Germany.\\
% ...
$^{94}$Department of Physics, Royal Holloway University of London, Egham; United Kingdom.\\
% ...
$^{102}$CPPM, Aix-Marseille Universit\'e, CNRS/IN2P3, Marseille; France.\\
$^{103}$Department of Physics, University of Massachusetts, Amherst MA; United States of America.\\
% ...
$^{128}$Homer L. Dodge Department of Physics and Astronomy, University of Oklahoma, Norman OK; United States of America.\\
% ...
$^{155}$Department of Physics and Astronomy, University of Sussex, Brighton; United Kingdom.\\
% ...
$^{160}$Raymond and Beverly Sackler School of Physics and Astronomy, Tel Aviv University, Tel Aviv; Israel.\\
% ...
$^{166}$Department of Physics, University of Toronto, Toronto ON; Canada.\\
% ...

\end{flushleft}


\end{document}